\documentstyle[12pt,aasms]{article}
\def\ecs{ergs cm$^{-2}$ s$^{-1}\ $}

\def\cts{cts s$^{-1}\ $ }

\def\lae{\mathrel{<\kern-1.0em\lower0.9ex\hbox{$\sim$}}}
\def\gae{\mathrel{>\kern-1.0em\lower0.9ex\hbox{$\sim$}}}

\received{:  January 15, 1999}
\accepted{:  March 29, 1999}

\slugcomment{Astrophysical Journal, in press}

\begin{document}

\title{A Medium Survey of the Hard X-Ray Sky with ASCA. II.: The Source's Broad Band X-Ray Spectral Properties}


\author
{Roberto Della Ceca\altaffilmark{1}, 
GianMarco Castelli\altaffilmark{1}, 
Valentina Braito\altaffilmark{1,2}, 
Ilaria Cagnoni\altaffilmark{3} and  
Tommaso Maccacaro\altaffilmark{1} }



\altaffiltext{1}{Osservatorio Astronomico di Brera, 
via Brera 28, 20121 Milano, Italy. E-mail: rdc@brera.mi.astro.it; 
tommaso@brera.mi.astro.it} 
\altaffiltext{2}{Universit\`a degli Studi di Milano, via Celoria 16, 
20133 Milano, Italy.}
 \altaffiltext{3}{International School for Advanced Studies, SISSA, via Beirut
2-4, 34014 Trieste, Italy.} 


\begin{abstract}

A complete sample of 60 serendipitous hard X-ray sources with flux in
the range $\sim 1 \times 10^{-13}$ \ecs to $\sim 4 \times 10^{-12}$
\ecs (2 - 10 keV), detected in 87 ASCA GIS2 images, was recently
presented in literature.  Using this sample it was possible to extend
the description of the 2-10 keV LogN($>$S)--LogS down to a flux limit
of $\sim 6\times 10^{-14}$ \ecs (the faintest detectable flux),
resolving about a quarter of the Cosmic X-ray Background.

In this paper we have combined the ASCA GIS2 and GIS3 data of
these sources to investigate their X-ray spectral properties using
the ``hardness" ratios and the ``stacked" spectra method.  Because of
the sample statistical representativeness, the results presented here,
that refer to the faintest hard X-ray sources that can be studied with
the current instrumentation, are relevant to the understanding of the
CXB and of the AGN unification scheme.

The ``stacked" spectra show that the average source's spectrum hardens
towards fainter fluxes: it changes  from an energy spectral index
$<\alpha_E> = 0.87 \pm 0.08$ for the 20 brightest sources (2-10 keV count
rate $\geq 3.9\times 10^{-3}$ \cts; the "Bright" sample) to
$<\alpha_E> = 0.36 \pm 0.14$ for the remaining 40 fainter sources (the
"Faint" sample).  The dividing line of $3.9 \times 10^{-3}$ \cts
corrispond to unabsorbed 2-10 keV fluxes in the range $\sim
5.4\times 10^{-13}$ to $\sim 3.1\times 10^{-13}$ \ecs for a source
described by a power law  model with energy spectral index between 0.0
and 2.0, respectively.
It thus seems that we are now beginning to detect those
sources that have the ``correct" spectral shape to be responsible for
the 2-10 keV CXB.

The hardness ratios analysis indicate that this flattening is due to a
population of sources with very hard spectra showing up in the Faint sample;
about half of the sources in this sample
require $\alpha_E \lae 0.5$ while only $\sim$ 10\% of the sources in
the brighter sample are consistent with an energy spectral index so flat.
A number of sources ($\sim 30\%$) in the Faint sample  
seem to be characterized by an apparently ``inverted'' X-ray
spectrum (i.e. $\alpha_E \lae 0.0$).  These objects are probably
extremely absorbed sources, as expected from the CXB synthesis models
based on the AGN Unification Scheme, if not a new population of very
hard serendipitous sources.

The broad band (0.7 - 10 keV) spectral properties of the selected
sources, as inferred from the hardness ratios diagram, seem to be more
complex than it is expected from a simple absorbed power-law model.  We
have thus investigated more complex models, in line with the AGNs
Unification Scheme, and we find that these latter models seem to be
able to explain the overall spectral properties of this sample; 
this result seems also suggested by the comparison of the hardness
ratio diagram of the serendipitous ASCA sources with that  obtained
using a sample of nearby and well known Seyfert 1 and Seyfert 2
galaxies observed with ASCA.

\end{abstract}


%
%

\section{Introduction}

In the last few years many efforts have been made to understand the origin 
of the cosmic X-ray background (CXB), discovered more than 35 years ago by 
Giacconi et al. (1962).
One of the competing hypotheses - the truly diffuse emission origin 
(see e.g. Guilbert and Fabian, 1986) -
has been rejected by the small deviation 
of the cosmic microwave background spectrum from a blackbody shape 
(Mather et al., 1994). Therefore only the alternative interpretation - the
discrete sources origin - is left.

Indeed, ROSAT deep surveys, reaching a source density of $\sim 1000$ 
deg$^{-2}$ at a limiting flux
of $10^{-15}$ \ecs (Hasinger et al., 1998; McHardy et al., 1998),
have already resolved the
majority (70$-$80 \%) of the soft ($E<2$ keV) CXB into discrete sources.
Spectroscopic observations (Shanks et al., 1991; Boyle et al.,
1993, 1994; McHardy et al., 1998; Schmidt et al., 1998) of the sources
with fluxes greater than $\sim 5 \times 10^{-15}$ \ecs have shown that the
majority ($50-80$ \%) of these objects are broad line AGN at $<z>
\sim 1.5$.  An important minority ($10-20$ \%) of ROSAT sources are
spectroscopically identified with X-ray luminous Narrow Emission Line
Galaxies (Griffiths et al., 1995, 1996; McHardy et al., 1998), whose
real physical nature (obscured AGN, starburst) is at the moment debated
in the literature (see e.g.  Schmidt et al., 1998). 
Since their average X-ray spectrum is harder than that of the broad line 
AGNs (Almaini et al., 1996; Romero-Colmenero et al., 1996)  and similar 
to that of the remaining unresolved CXB, these objects could also be 
substantial contributors to the CXB at higher energies. 
About 10\% of the
ROSAT sources are identified with clusters of galaxies (see e.g. Rosati et
al.,1998).  Thus, it is clear that the ROSAT satellite has been
successful in resolving almost all the soft CXB into discrete sources.
Furthermore, 
optical observations of the faint ROSAT sources has lead to the 
understanding of the 
physical nature of the objects contributing to it.

On the contrary, at harder energies, closer to
where the bulk of the CXB resides, the origin of the CXB
is still matter of debate.
Before the ASCA and Beppo-SAX satellites, which carry the first
long lived imaging instruments in the 2-10 keV energy band, surveys in
this energy range were made using passively collimated X-ray detectors
that, because of their limited spatial resolution, allowed the
identification of the brightest X-ray sources only, which represent a
very  small fraction ($< 5\%$) of the CXB (Piccinotti et al., 1982).
 The so called ``spectral paradox"
further complicates the situation: none of the single classes of
X-ray emitters in the Piccinotti et al., (1982) sample is characterized
by an energy spectral distribution similar to that of the CXB.  Due to
the lack of faint, large and complete samples of X-ray sources selected
in this energy range, the contribution of the different classes of
sources to the hard CXB was evaluated through population
synthesis models and different classes of X-ray sources were proposed
as the major contributors by a number of authors
(e.g. starburst
galaxies, absorbed AGN, reflection dominated AGN,
see e.g. Griffiths and Padovani, 1990; Madau, Ghisellini and Fabian, 1994; 
Comastri et al., 1995; Zdziarski et al., 1993).
Recent results from ASCA and Beppo-SAX observations of individual objects
and/or medium-deep survey programs
(Bassani et al., 1999; Maiolino et al., 1998a; Turner et al., 1998; Boyle et al., 
1998a; 1998b; Akiyama et al., 1998)
seem to
favor the strongly absorbed AGN hypothesis but deeper investigations
are still needed to confirm this scenario.

At the {\it Osservatorio Astronomico di Brera}, a serendipitous search
for hard (2-10 keV band) X-ray sources using data from the GIS2
instrument onboard the ASCA satellite is in progress (Cagnoni, Della
Ceca and Maccacaro 1998, hereafter Paper I; Della Ceca et al.  1999, in
preparation) with the aim of extending to faint fluxes the census of
the X-ray sources shining in the hard X-ray sky.  The strategy of the
survey, the images and sources selection criteria and the definition of
the sky coverage are discussed in Cagnoni, Della Ceca and Maccacaro
(1998).

In Paper I, a first sample of 60 serendipitous X-ray sources, detected 
in  87 GIS2 images at high  
galactic latitude ($|b| > 20^o$) covering $\sim 21$ square deg.
was presented. 
This sample has allowed the authors to extend the description of the
number-counts relationship down to a flux limit of $\sim 6\times
10^{-14}$ \ecs (the faintest detectable flux), 
resolving {\bf directly} about 27\% of the (2-10 keV) CXB.


Here we study the spectral properties of the 60 ASCA sources listed in
Paper I, combining GIS2 and GIS3 data. We have carried out both an
analysis of the ``stacked" spectra of the sources in order to
investigate the variation of the source's average spectral properties
as a function of the flux, and a ``hardness-ratio" (HR) analysis of the
single sources.  This latter method, which is equivalent to the
``color-color" analysis largely used at optical wavelengths, is
particularly appropriate when dealing with  sources detected at a low
signal-to-noise ratio (e.g.  Maccacaro et al., 1988; Netzer, Turner and
George, 1994).  We have defined two independent HR and we have compared
the position of the sources in the HR diagram with a grid of
theoretical spectral models which are found to describe the X-ray
properties of known classes of X-ray emitters.

The paper is organized as follows.  In section 2 we present the sample
and we define the ``Faint'' and the ``Bright'' subsamples. In section 3 we
present the data, we discuss the data analysis and we define the two HR
used.  In section 4 we  report the results of the stacked spectra and
of the HR analysis and we compare them with those expected from simple
spectral models and with the CXB spectra.  Summary and conclusions are
presented in section 5.
 
\vfill
\eject

\section{Definition of the ``Faint" and ``Bright" Subsamples}

The basic data on the 60 X-ray sources used in this paper are reported in 
Table 2 of Paper I.

To investigate if the spectral properties of the sources depend on
their brightness we have defined two subsamples according to the
``corrected'' 2-10 keV count rate (hereafter CCR)  
\footnote{Due to the vignetting of the XRT and the PSF, a source with a
given flux will yield an observed count rate depending on the position
of the source in the field of view.  With ``corrected count rate" we
mean the count rate that the source would have had if  observed at some
reference position in the field of view and   whitin a given  extraction region.
The definition of the source extraction region and of the reference position in the 
field of view, used to determine the corrected CCR, are discussed in section 3.2.
The CCR used here can be obtained dividing the
unabsorbed 2-10 keV flux reported in Paper I by the count rate to
flux conversion factor of 1cnt/s (2-10) = $11.46 \times 10^{-11}$ \ecs.
This value is appropriate for a power law
model with energy spectral index of 0.7, filtered by a Galactic
absorbing column density of $3 \times 10^{20}$ cm$^{-2}$.}
.

The 20 brightest sources (CCR $\geq 3.9 \times 10^{-3}$ \cts) define the
``Bright" subsample, while the remaining 40 sources define the ``Faint"
subsample.
The dividing line of $3.9 \times 10^{-3}$ \cts corrisponds to an 
unabsorbed 2 - 10 keV flux of 
$\sim 5.4\times 10^{-13}$ or $\sim 3.1\times 10^{-13}$ \ecs 
for a source described by a power law  model with energy spectral 
index of 0.0 and 2.0, respectively absorbed by  a
Galactic column density of $3 \times 10^{20}$ cm$^{-2}$.
We note that the numbers reported above are a very weak function of the
Galactic absorbing column density along the line of sight (which ranges
from $10^{20}$ cm$^{-2}$ to $9 \times 10^{20}$ cm$^{-2}$ for the
present sample).

We prefer to use the CCR instead of the flux because the CCR is (once
the corrections due to the vignetting and the PSF have been applied) an
{\it observed} quantity and it is independent on the spectral
properties of the source; on the contrary the flux is model
dependent.  In first approximation fainter CCR corresponds to fainter
sources.
 
\section{Data analysis}

All the ASCA images used in Paper I are now in ``REV 2"
Processing status 
(see http://heasarc.gsfc.nasa.gov/docs/asca/ascarev2.html), thus
in this paper we have used this new revision of the data.
Furthermore, in order to improve the statistics, we have combined
\footnote{This combination was possible for all the sources but a0447-0627, 
a0506-3726, a0506-3742 and a0721+7111. For these 4 sources we have used 
only the GIS2 data.}
data from the GIS2 and GIS3 instruments as explained below.

Data preparation has been done using version 1.3 of the XSELECT
software package and version 4.0 of FTOOLS (supplied by the HEASARC at
the Goddard Space Flight Center).  Good time intervals were selected
applying the ``Standard REV 2 Screening" criteria  (as reported in
chapter 5 of the ASCA Data Reduction Guide, rev 2.0), with the only
exception of having used a magnetic cutoff rigidity threshold of 6 GeV
c$^{-1}$ (as done in Paper I).  HIGH, MEDIUM and LOW bit rate data were
combined together.  Spectral analysis (see below) has been performed
using version 9.0 of the XSPEC software package.  We use the detector
Redistribution Matrix Files (RMF) gis2v4\_0.rmf and gis3v4\_0.rmf.

\vfill
\eject

\subsection{``Stacked" spectra}

For each source and for the GIS2 and GIS3 data sets,
total counts (source + background) were extracted from
a circular extraction region of 2 arcmin radius around the source centroid.
Background counts  were taken from two
circular uncontaminated regions of 3.5 arcmin radius, close to the source,
or symmetrically located with respect of the center of the image.  Source and 
background data were
extracted in the ``Pulse Invariant" (PI) energy channels, which have
been corrected for spatial and temporal variations of the detector
gain.  The Ancillary Response File (ARF) relative to each source was
created with version 2.72 of the FTOOLS task ASCAARF
\footnote{ The task ASCAARF (which is part of the FTOOLS software
package) is able to produce a position-dependent PSF-corrected
effective area, A(E,x,y,d), of the (XRT + GIS2 or XRT + GIS3) combination. The
inputs of this task are the position {\it x,y} (in detector
coordinates) and the dimension {\it d} of the source's extraction
region. }
at the
location of the individual sources in the detectors.

For each source we have then produced a combined GIS spectrum 
(adding GIS2 and GIS3 data) and the corresponding
background and response matrix files, following the
recipe given in the ASCA Data Reduction (rev 2.0) Guide (see section
8.9.2 and 8.9.3 and reference therein).
Finally, we have produced the combined
spectrum of 
a) the 20 sources belonging to the Bright Sample and 
b) the 40 sources of the Faint Sample. 
We note that each object contributes to the stacked counts at most for
6\% in the case of the Faint Sample and at most for 25\% in
the case of the Bright Sample.

In the spectral analysis, being interested in comparing these stacked
spectra with that of the hard CXB, we have considered only the counts
in the 2.0 - 10.0 keV energy range. The total net counts in the Bright
and Faint sample are about 3400 and 2900 respectively.  The stacked
spectra were rebinned to give at least 50 total counts per bin.

\subsection{Hardness Ratios}

For each source and for the GIS2 and GIS3 data set, source plus
background counts were extracted in three energy bands:  0.7-2.0  keV
(S band), 2.0-4.0  keV (M band) and 4.0-10.0 keV (H band).  The S, M
and H spectral regions were selected so as to have similar statistics
in each band for the majority of the sources.  The background counts
have been evaluated by using the two background regions considered
above; first, we have normalized the background counts to the source
extraction region and then we have averaged them.  Net counts in S, M
and H have been then obtained for each source by subtracting the
corresponding S,M and H normalized background counts from the total ones.


To combine the GIS2 and GIS3 data for each source and to
compare our results with those expected from simple models, the net counts obtained
 must be corrected for the position dependent sensitivity
of the GIS detectors (see the discussion in section 2.3 of Paper I).
In particular we must:
a) define a source extraction region and a reference position in the GIS2 or
GIS3 field;
b) re-normalize the S,M,H net counts from each source to this region (since 
sources are detected in different locations of the GIS2 or GIS3 field of view);
c) perform the simulations for simple models by using the effective area of 
this region. We will now discuss these points in turn.

As reference region we have used a source extraction region of 2 arcmin
radius at the position x=137, y=116 for the (XRT + GIS2) combination.
Using ASCAARF we have produced the effective area values at the
position of each source detected in the GIS2 (GIS3) detector.  Using
these effective area values and through spectral simulations with XSPEC
we have derived the correction factors to be applied to each source in
the S, M and H band.  
For each source we have then applied these correction factors to
the net counts of GIS2 and GIS3 separately.  
Finally, we
have combined the GIS2 and GIS3 corrected net counts for each source.
In summary, using this procedure we have first re-normalized the GIS2
and GIS3 data for each source to the reference region separately and
then combined them.
We note that the applied method is very similar to the
``flat field" procedure normally used in the analysis of optical
imaging and spectroscopic data.

Using the corrected net counts in the S, M and H band we have then computed 
for each source two ``hardness ratios" defined in the following way: 
$$HR1 = {M-S \over M+S}\ \ \ \ \ \ \  HR2 = {H-M \over H+M}$$

The 68\% error bars on HR1 and HR2 have been obtained via Monte Carlo simulations  
using the total counts, the background counts and the 
correction factors relative to each source.

Similarly, HR1 and HR2 values expected from simple spectral models (see
section 4) have been obtained with XSPEC using the effective area of
the reference region.

\section{Results}
 
\subsection{The Hard Energy Range and the CXB}

In this section we discuss the hard (2-10 keV) X-ray average spectrum of the
present sample and we compare it with that of the CXB; to this end we
use the HR2 values and the ``stacked" spectra introduced in section 3.

In figure 1, for all sources, we plot the HR2 value versus the GIS2 CCR; 
the filled squares represent the sources detected with a
signal to noise ratio greater than 4.0 while the open squares represent
the sources detected with a signal to noise ratio between 3.5 and 4.0.
The HR2 values are then compared with those expected from a non
absorbed power-law model with energy spectral index $\alpha_E$ ($f_X
\propto E^{-\alpha_E}$) ranging from $-$1.0 to 2.0.

Figure 1 clearly shows a broadening of the HR2 distribution going to
fainter CCR; furthermore a flattening of the mean spectrum with
decreasing fluxes is evident.  A similar broadening and flattening of
the HR2 distribution is still detected if only the 40 sources (18
Bright and 22 Faint) detected with a signal to noise ratio greater than
4.0 are used, showing that this result is not due to the sources near
the detection threshold limit.

It is worth noting the presence of many sources which seem to be
characterized by a very flat 2-10 keV spectrum with $\alpha_E \leq 0.5$
and of a number of sources with ``inverted" spectra (i.e.  $\alpha_E \leq
0.0$).  This is particularly evident in the Faint Sample where about
half of the sources seem to be described by  $\alpha_E \leq 0.5$ and
about 30\% by ``inverted" spectra. These latter objects could represent
a new population of very hard serendipitous sources or, alternatively,
a population of very absorbed sources as expected from the CXB
synthesis models based on the AGN Unification Scheme.

We have checked whether the observed hardening of the mean spectral
index can be attributed to a spectral bias in the source's selection.
Since sources with the same flux but different spectrum will deposit a
different number of counts in the detector, it is evident that, as one
approaches the flux limit of a survey, sources with a {\it favourable}
spectrum will be detected (and thus included in the sample), while
sources with an {\it unfavourable} spectrum will become increasingly
under-represented (see Zamorani et al., 1988 for a discussion of this
effect). However, in the case of the ASCA/GIS, this selection effect
favours the detection of steep sources, thus giving further support to
the reality of our findings
\footnote{As an example, in a 50.000 s observation, a source with 2-10
keV flux = $5 \times 10^{-13}$ \ecs characterized by an absorbed
power-law spectrum with $\alpha_E = 2.0$ and $N_H = 3\times10^{20}$
will deposit (at the reference position and inside a region of 2 arcmin
radius) $\sim 300$ (2-10) keV counts while a source with the same flux
but power-law spectrum with $\alpha_E = 0.0$ and same $N_H$ will
deposit $\sim 180$ counts.}.

We note that a population of very hard X-ray sources has been recently
suggested also by Giommi et al., 1998 in order to explain the spectral
properties of the faint BeppoSAX sources and to reconcile the 2-10 keV
LogN($>$S)--LogS with the 5-10 keV LogN($>$S)--LogS obtained from
BeppoSAX data.

To further investigate the flattening of the source's mean spectral
index we have used the ``stacked" spectra introduced in section 3.1.

In Table 1 we report the results of the power-law fits to the stacked
spectra of the Faint and the Bright sample; the unfolded spectra
of the two samples are shown in figure 2 (Bright sample: open squares; 
Faint sample: filled squares).

As can be seen from the $\chi^{2}_{\nu}$ values reported in Table 1
and from the spectra shown in figure 2, a simple power-law spectral
model represents a good description of the stacked spectra of the two
samples in the 2-10 keV energy range.  The $N_H$ values corresponding
to the mean line of sight Galactic absorption for the two independent
samples have been used in the fits.  However, given the energy range of
interest (2.0 - 10.0 keV), the spectra are not significantly affected
by the Galactic $N_H$ value.  Consistent results are obtained if we use
either the lowest or highest ($0.7 \times 10^{20}$ cm$^{-2}$ or $9.06
\times 10^{20}$ cm$^{-2}$ respectively) Galactic $N_H$ value sampled in
the present survey.

It is worth noting that the unfolded spectra of the Faint and the Bright
sample
reported in figure 2 does not show any compelling evidence of emission lines.
A strong emission line should be expected if, for example, sources with
a strong Iron emission line at 6.4 keV, contributing
in a substantial way to the CXB, were strongly clustered at some particular
redshift. This subject has been already discussed by Gilli et
al., 1999, reaching the conclusion that the maximum
contribution of the Iron line to the CXB is less than few percent.
The unfolded spectra shown in figure 2 confirm their results.

In Figure 3 the results obtained here are compared with those obtained
using data from other satellites or from other ASCA medium-deep survey
programs.  The best-fit energy index of the Bright sample ($<\alpha_E>
= 0.87 \pm 0.08$) is in good agreement with the mean spectral
properties of the objects in the Ginga and HEAO1 A-2 sample and with
the mean spectral properties of the broad line AGNs detected by ROSAT
(Almaini et al., 1996; Romero-Colmenero et al., 1996).  The Faint
sample is best described by  $<\alpha_E> = 0.36 \pm 0.14$; this is
consistent with other ASCA results (Ueda et al., 1998) and with the
spectra of the CXB in the 2-10 keV energy range (Marshall et al., 1980;
Gendreau et al., 1995).  For the purpose of figure 3 we have
used a count rate to flux conversion factor adeguate for a power law
spectral model with $\alpha_E = 0.36$ (Faint sample) or $\alpha_E =
0.87$ (Bright sample).

We have evaluated the influence that the sources with the hardest energy 
distribution have on the combined spectra. If we exclude the 6 sources
with HR2$>$0.2 (5 from the Faint sample and 1 from the Bright sample)
we find that the combined spectra are still significantly different,
being described by a power-law model with $<\alpha_E> = 0.53 \pm 0.14$ 
and $<\alpha_E> = 1.04 \pm 0.10$ respectively.

Finally, in the case of the Bright sample two sources contribute about 35\%
of the total counts; if we exclude these two objects the
remaining stacked spectrum is described by a power-law model with
$<\alpha_E> = 1.00 \pm 0.16$, showing that the inclusion or the
exclusion of these two objects do not change
any af our results.  Note that in the case of the Faint sample each
object contributes to the stacked counts at most for 6\% or less.

These results clearly show that: 
a) we have detected a flattening of the source's mean spectral 
properties toward fainter fluxes and 
b) we are beginning to detect those X-ray sources having a combined
X-ray spectra, consistent with that of the 2 - 10 keV CXB.

\subsection{The Broad Band Spectral Properties and the AGN Unification Scheme}

In this section we do not intend to derive specific spectral properties
and/or parameters for each source; the limited statistics and the
complexity of AGNs broad band X-ray spectra (see Mushotzky, Done
and Pounds, 1993 for a review of the subject) prevent us from doing  so.
Rather we regard this sample as representative of the hard X-ray sky
and we try to investigate if the currently popular CXB synthesis models
based on the AGNs Unification Scheme can describe the overall spectral
properties of the ASCA sample as inferred from the hardness ratios.
According to the AGNs Unification model for the 
synthesis of the CXB, a population of unabsorbed (Type 1:
$N_{H} \lae 10^{22}$ cm$^{-2}$) and absorbed (Type 2: $N_{H} \gae 10^{22}$ 
cm$^{-2}$) AGNs can reproduce the shape and intensity of the
CXB from several keV to $\sim$ 100 keV (see Madau, Ghisellini and
Fabian, 1994; Comastri et al., 1995; Maiolino et al., 1998b).
Because $\sim 90\%$ of the ASCA sources in this sample 
\footnote{Up to now 12 sources have been spectroscopically identified.
The optical breakdown is the following: 1 star, 2 cluster of galaxies,
7 Broad Line  AGNs and 2 Narrow Line AGNs.  However we stress that this
small sample of identified objects is probably not representative of
the whole population.}
are expected to be AGNs (see Paper I), in the following we will consider this sample 
as being well approximated by a population of (Type 1 + Type 2) AGNs  
with flux above $\sim 1 \times 10^{-13}$ \ecs.

In figure 4 we show (open squares) the position of the sources in the
``Hardness Ratio" diagram for the Faint (panel a) and Bright (panel b)
sample; in figure 4c we have reported the 12 sources spectroscopically
identified so far (see footnote 8).  Also reported in figure 4a,b,c
(solid lines) are the loci expected from X-ray spectra described by an
absorbed power law model.  The energy index of the power law  ranges
from $-$1.0 up to 2.0, while the  absorbing column densities ranges
from  $10^{20}$ cm$^{-2}$ up to $10^{24}$  cm$^{-2}$ (see figure 4d); 
the absorption has been assumed to be at zero redshift (i.e. the
sources are supposed to be in the local universe).

Figure 4 clearly shows how the ``Hardness Ratio" diagram, combined with
spectral simulations, can be used to obtain information on the sources'
X-ray spectral properties, as well as the power of using two hardness
ratios to investigate the broad band spectral properties of the
sources.  If only one ratio is known, say HR2 for example, a source
with HR2 $\sim 0-0.1$ could be described  by an absorbed  power law
with either energy index of 0.0 and $N_H$ = 10$^{20}$ cm$^{-2}$ or energy
index of $\sim $ 2.0 and $N_H$ =  10$^{23}$ cm$^{-2}$.  The use of both
HR1 and HR2 allows the ambiguity to be solved.

We note that the hardness ratios computed as described in section 3.2
have not be corrected for the different Galactic absorbing column
density along the line of sight of each source.  For the present sample
this ranges from $10^{20}$ cm$^{-2}$ to $9 \times 10^{20}$ cm$^{-2}$;
figure 4 clearly show that this correction is insignificant.

One of the most striking features of figure 4 is the large spread in HR1
and HR2 displayed by the ASCA sources and the departure from the loci
of absorbed, single power law spectra.  This implies that the broad
band (0.7 - 10 keV) spectrum of the sources is more complex than the
simple model of an absorbed power law. In particular the ASCA sources
located on the left side of the line representing power laws with $N_H
\sim$ 10$^{20}$ cm$^{-2}$ (see figure 4d) are not explained within the
absorbed power-law model.  A similar result is obtained even if the
absorbing material is assumed to be at higher redshift e.g. z = 1 or 2.

We note that the object in figure 4b (or 4c) characterized by HR1
$\sim$ 0.0 and  HR2 $\sim 0.71$  (a1800+6638; its X-ray flux is $\sim
6.5 \times 10^{-13}$ \ecs) is spectroscopically identified with the
Seyfert 2 galaxy NGC 6552. The X-ray spectrum of this object (Reynolds
et al., 1994 and Fukazawa et al., 1994) is consistent with a model
composed of a narrow Gaussian line plus an empirical ``leaky-absorber"
continuum; the latter is composed by an absorbed power-law and a non
absorbed power law having a common photon index.  Fukazawa et al.
(1994) found that the NGC 6552 spectrum requires a photon index of
$\sim 1.4$, an absorbing column density of $\sim 6\times 10^{23}$
cm$^{-2}$, an uncovered fraction of $\sim 2\%$ and a narrow Gaussian
line (consistent with the $K_{\alpha}$ iron emission line at 6.4 keV)
having an equivalent width of $\sim $  0.9 keV.

That the simple absorbed power law model is unable to explain the
scatter in the hardness ratios is not surprising given the
observational evidence  that the broad band X-ray spectra of Type 1
and, even more, of Type 2 AGN are complex and affected by several
parameters like, for instance, the viewing angle and the torus
thickness (see e.g. Turner et al., 1997; Turner et al., 1998;
Maiolino et al., 1998a; Bassani et al., 1999).  In particular, for Type
2 AGNs we could have, as a function of the absorbing column density,
one of the following three cases:

1) $10^{22} \lae N_H \lae 5\times 10^{23}$ cm$^{-2}$.  The hard X-ray
continuum above a few keV is dominated by the directly-viewed
component making the source nucleus visible to the observer and the
column density measurable; the reflected/scattered component is
starting to become relevant in the soft energy range.  In this case
the observed spectrum is that of an absorbed Type 1 AGNs with some
extra flux at low energies;

2) $5\times 10^{23} \lae N_H \lae 10^{25}$ cm$^{-2}$. In this case 
both the directly-viewed component and the reflected/scattered 
component are observed and the resulting spectrum becomes
very complex.

3) $N_H \gae 10^{25}$ cm$^{-2}$. The torus is very thick. The
continuum source is blocked from direct view up to several tens of
keV and the observed spectrum is dominated by the scattered/reflected component.  
In this case the observed continuum is that of a Type 1 AGNs (in the
case of scattering by warm material near the nucleus) or that of a
Compton reflected spectrum (in the case of cold reflection from the
torus).

Given this spectral complexity we have tested the Hardness Ratios 
plot against the following simplified AGNs spectral model composed of:

a) an absorbed power-law spectrum. This component represents the continuum
source, i.e. for an absorbing column density of about $10^{20}$ cm$^{-2}$
this represents the ``zero" order continua of a Type 1 object;

b) a non absorbed power-law component, characterized by the same energy
spectral index of the absorbed power-law and by a
normalization in the range 1-10\% of that of the absorbed power-law
component.  This component represents the scattered fraction (by warm
material) of the nuclear emission along the line of sight;  
theoretical and observational evidence (see Turner et al., 1997 and
reference therein) suggest that this scattered fraction is in the range
1-10\%;

c) a narrow emission line at 6.4 keV, having an equivalent width of
230 eV for low values of $N_H$ ($\sim 10^{20}$ cm$^{-2}$).  This
component represents  the mean equivalent width of the Fe K$_{\alpha}$
emission line in Seyfert 1 galaxies as determined by Nandra et al.,
1997 using ASCA data.

In summary, the AGN spectral model used is: 
$$S(E) = E^{-\alpha_{E}} e^{-\sigma_{E} \times N_{H}} +
         K E^{-\alpha_{E}} + Fe_{6.4 keV}$$
where E is the photon energy, $\alpha_{E}$ is the energy spectral
index, $\sigma_{E}$ is the energy dependent absorbing cross-section,
$N_{H}$ is the absorbing column density along the line of sight to the
nucleus, K is the scattered fraction and $Fe_{6.4 keV}$ is the Iron
narrow emission line at 6.4 keV.
The above simplified AGNs spectral model is equivalent to the model
used, by Fukazawa et al., 1994 in the case of NGC 6552 and, by several
authors to describe the ``first order" X-ray spectra of
Seyfert 2 galaxies (see e.g. Turner et al., 1997).

We have tested this model as a function of $\alpha_{E}$, the absorbing
column density $N_{H}$ and the scattered fraction K.  Figure 5 shows,
as an example, the results of the spectral simulations in the case of
local sources ($z\sim 0.0$) with $\alpha_{E} = 1$ and K = 0.01 (filled
triangles) or K = 0.1 (filled circles).  The input spectral models for
the case K = 0.01 and Log$N_H$ = 20, 22, 23, 24 and 24.5  are also shown.

We would like to stress that using this simplified AGNs spectral model
we can go, in a continuous way, from a  ``first order" Type 1 spectra to a
``first order" Type 2 spectra only by changing the $N_H$ parameter.  In
this respect we note that the equivalent width of the Iron narrow
emission line at 6.4 keV (which, as we said, is fixed to 230 eV for
$N_H \sim 10^{20}$ cm$^{-2}$) is  $\sim 2$ keV when $N_H \sim
3\times 10^{24}$ cm$^{-2}$; this latter value is very similar to that
measured in many Seyfert 2 galaxies characterized by a similar value of
the absorbing column density (Bassani et al., 1999).

As anticipated at the beginning of this section some qualitative
conclusions can be drawn from the results reported in Figure 5.  First
of all it appears  that in the context of this simplified AGN
spectral model, and for very high  $N_H$ values,  we are able to
explain the hardness ratios of the objects located on the left side of
the dashed line representing unabsorbed power laws.  Second, the
results reported in Figure 5 seem to indicate that many of the latter
objects could be characterized by a very high absorbing column density;
if this indication is confirmed by deeper investigation with XMM
and AXAF, then the number of Compton ``thick" systems could be
significantly higher than previously estimated.  In the meantime we
note that this result is consistent with recent findings obtained by
Maiolino et al., 1998a, Bassani et al., 1999 and 
Risaliti, Maiolino and Salvati, 1999 by studying the fraction
of  Compton ``thick" systems in a sample of Seyfert 2 galaxies observed
with BeppoSAX and/or with ASCA.

Finally, to investigate the Hardness Ratio plot {\it in a model
independent way}, we have used a comparison sample of nearby (z$<$1)
Seyfert 1 and Seyfert 2 galaxies pointed at with ASCA. This sample was
selected from a data set of about 300 ASCA pointings that were treated
in the same manner (see section 2) and that are used to extend the
survey presented in Cagnoni, Della Ceca and Maccacaro (1998).  Amongst
this restricted ASCA data set we have considered those observations
pointed on nearby Seyfert 1 and Seyfert 2 galaxies; furthermore, we
have considered only the Seyfert 2 galaxies also reported in Table 1 of
Bassani et al.,1999 in order to have an uniform data set on their
Compton thickness.
The comparison sample is composed of 13
Seyfert 1, 15 Compton thin Seyfert 2 and 7 Compton thick Seyfert 2
\footnote{The Seyfert 1 are: NGC 1097, MKN 1040, RE 1034+39, MKN 231, MKN 205, 
3C 445, I ZW 1, 2E1615+0611, MKN 478, EXO055620-3820.2, MKN 507, NGC 985, TON 180. 
          The Compton thin Seyfert 2 are:  NGC 3079, NGC 4258, NGC 3147, NGC 5252, 
MKN 463e, NGC 1808, NGC 2992, NGC 1365, NGC 6251, MKN 273, IRAS05189-2524, NGC 1672, 
PKS B1319-164, MKN 348, NGC 7172.
          The Compton thick Seyfert 2 are: MKN 3, NGC 6240, NGC 1667, NGC 4968, NGC 1386, 
MKN 477, NGC 5135.}.
The targets were analyzed in the same way as the 
serendipitous sources, combining the GIS2 and GIS3 data 
(see section 3.2). 

The results are shown in figure 6.  The Seyfert 1 galaxies are strongly
clustered around the locii representing low $N_H$ values; on the
contrary the Seyfert 2 galaxies are characterized by a very large
spread in their HR1, HR2 values.  In particular many of the Seyfert 2
are located on the left side of the line representing power laws with
$N_H \sim 10^{20}$ cm$^{-2}$; about a half of these sources have been
classified as Compton thick system from Bassani et al., 1999.  We are
tempted to suggest that also the other half are Compton thick objects,
whose nature is still unrevealed because of the presently poor data
quality.  XMM and/or AXAF observations are needed to confirm this
suggestion.  However, whatever their properties (Compton Thin or Thick)
are, the comparison of figures 6 and figures 4a,b strongly suggest
that at least some of the serendipitous ASCA sources located on the
left side of the dashed line connecting the $N_H=10^{20}$ values could
be Type 2 AGNs.

\section{Summary and Conclusion}

In this paper we have used ASCA GIS2 and GIS3 data for a complete
and well defined sample of 60 hard (2-10 keV) selected sources to study
the spectral properties of  X-ray sources down to a flux limit 
of $\sim 10^{-13}$ \ecs.
To investigate if the spectral properties of the sources depend on
their brightness we have defined two subsamples according to the
``corrected'' 2-10 keV count rate; the Bright sample is defined
by the 20 sources with CCR $\geq 3.9\times 10^{-3}$ \cts, while the
Faint sample is defined by the 40 fainter sources.
The dividing line of $3.9 \times 10^{-3}$ \cts
corrispond to unabsorbed 2-10 keV fluxes in the range $\sim
5.4\times 10^{-13}$ to $\sim 3.1\times 10^{-13}$ \ecs for a source
described by a power law  model with energy spectral index between 0.0
and 2.0, respectively.

The main results of this investigation are the following:

a) the average (2-10 keV) source's spectrum hardens towards fainter
fluxes.  The ``stacked" spectra of the sources in the Bright sample
is well described by a power-law model with
an energy spectral index of $<\alpha_E> = 0.87\pm 0.08$, while the
``stacked" spectra of the sources in the Faint sample
require $<\alpha_E> = 0.36 \pm 0.14$; this means that we are beginning
to detect those
sources having a combined X-ray spectrum consistent with that of the
2-10 keV CXB.

b) the hardness ratios analysis shows that this flattening is due to
the appearance of sources with very hard spectra in the Faint sample. 
About a half of the sources in the latter sample
require $\alpha_E \lae 0.5$ while only 10\%
of the brighter sources are consistent with an energy spectral index so
flat.  Furthermore about 30\% of the sources in the Faint sample
seem to be characterized by ``inverted'' ($\alpha_E \lae
0.0$) 2-10 keV X-ray spectra.  These latter objects could represent a
new population of very hard serendipitous sources or, alternatively, a
population of very absorbed sources as expected from the CXB synthesis
models based on the AGN Unification Scheme (see below).

c) a simple absorbed power-law model is unable to explain the broad
band (0.7 - 10 keV) spectral properties of the sources, as inferred
from the Hardness Ratios diagram.  
X-ray spectral
models based on the AGNs Unification Scheme seem to be able to explain
the overall spectral properties of this sample.  This seems also
suggested by
the comparison of our results with those obtained using a 
sample of nearby and well known Seyfert 1 and Seyfert 2 galaxies
observed with ASCA.

Acknowledgments

We are grateful to L.Bassani for stimulating discussions on the
comparison sample of Seyfert galaxies, A.Wolter for a carefull reading
of the manuscript and the anonymous referee for useful  comments.  G.C.
acknowledge financial support from the ``Fondazione CARIPLO".  This
work received partial financial support from the Italian Ministry for
University and Research (MURST)  under grant Cofin98-02-32.  This
research has made use of the NASA/IPAC extragalactic database (NED),
which is operated by the Jet Propulsion Laboratory, Caltech, under
contract with the National Aeronautics and Space Administration.  We
thank all the members of the ASCA team who operate  the satellite and
maintain the software data analysis and the archive.

%
%
%
%

\newpage
\begin{center}
Table 1\\
\vskip 0.5truecm
\begin{tabular}{ccccrr} \hline \hline
Sample & Objects & Net counts & $\alpha_E$ & $N_{H Gal}$ & $\chi^2_{\nu}$/$\nu$\\ \cline{1-6} \hline
 & & $(2-10)$ keV & & $(10^{20}cm^{-2})$ &\\ \hline
Faint & 40 & $\sim 2900$ & $0.36\pm0.14$ & 2.75 & 1.09/105\\ 
Bright & 20 & $\sim 3400$ & $0.87\pm0.08$ & 3.66 & 1.06/82\\ \hline  
\end{tabular}
\end {center}

%

\clearpage
\section{Figure Captions:}

Figure 1: HR2 value versus the "corrected" 2 - 10 keV count rate
for all the serendipitous
ASCA sources belonging to the Cagnoni, Della Ceca and Maccacaro, 1998
sample. The filled squares represent the sources detected with a signal
to noise ratio greater than 4.0 while the open squares represent the
sources detected with a signal to noise ratio between 3.5 and 4.0.  The
dotted lines represent the expected HR2 values for a non absorbed
power-law model with an energy spectral index as indicated in the
figure.  The dashed line represent the dividing line between the Faint
and the Bright sample.

Figure 2: The unfolded stacked ASCA spectrum of the Faint (filled
squares) and the Bright (open squares) sample.
Note that this spectra are shown in arbitrary units.

Figure 3: Comparison of the results obtained from the stacked spectra
of the Faint and the Bright sample with those obtained using data from
other satellites or from other ASCA medium-deep survey programs
(adapted from figure 3 of Ueda et al., 1998).

Figure 4:
a)  ``Hardness Ratio" distribution  for the sources in the Faint
sample.  The grid  represents the locii expected from 
absorbed power law spectra (values are shown in panel d);
b) As panel a but for the Bright sample;
c) As panel a but for the identified sources. 
Symbols are as follows:  star: star; filled triangles:
clusters of galaxies; filled squares:  Narrow Line AGNs; open squares:
Broad Line AGNs;
d) Expected HR1 and HR2 values as a function of the absorbing column density and 
of the power law  energy index of the spectrum.
The absorption is assumed to be at zero redshift.  

Figure 5: The open squares represent the position of the sources in the
``Hardness Ratio" diagram (Faint + Bright Sample).  The solid lines are
the locii expected from the simplified AGN spectral model discussed in
the text (see section 4.2). We have  highlighted with filled dots or
filled triangles the absorbing column densities of Log $N_H$ = 20,
21.5, 22, 22.5, 23, 23.5, 24, 24.5 (from left to right); the absorption
has been assumed to be at zero redshift (i.e. the sources are supposed to be
in the local universe).  The filled triangles represent the model
relative to an energy spectral index equal to 1.0 and scattered
fraction of 1\%, while the filled dots represent the model relative to
an energy spectral index equal to 1.0 and scattered fraction of 10\%.
The dashed line represents the locus expected from unabsorbed power
laws ($N_H \sim 10^{20}$ cm$^{-2}$) and energy spectral index ranging
from  $-$1.0 up to 2.0.  We also show the input spectral
models for the case of K=10\% and LogN$_H$ = 20, 22, 23, 24, and 24.5.

Figure 6: As figure 4a but for the comparison sample of Seyfert 1
and Seyfert 2 galaxies. 
Filled triangles: Seyfert 1; Open squares: Compton Thick Seyfert 2; 
Filled Squares: Compton Thin Seyfert 2 (see section 4.2).

\clearpage
\begin{figure}
\figurenum{1}
\plotone{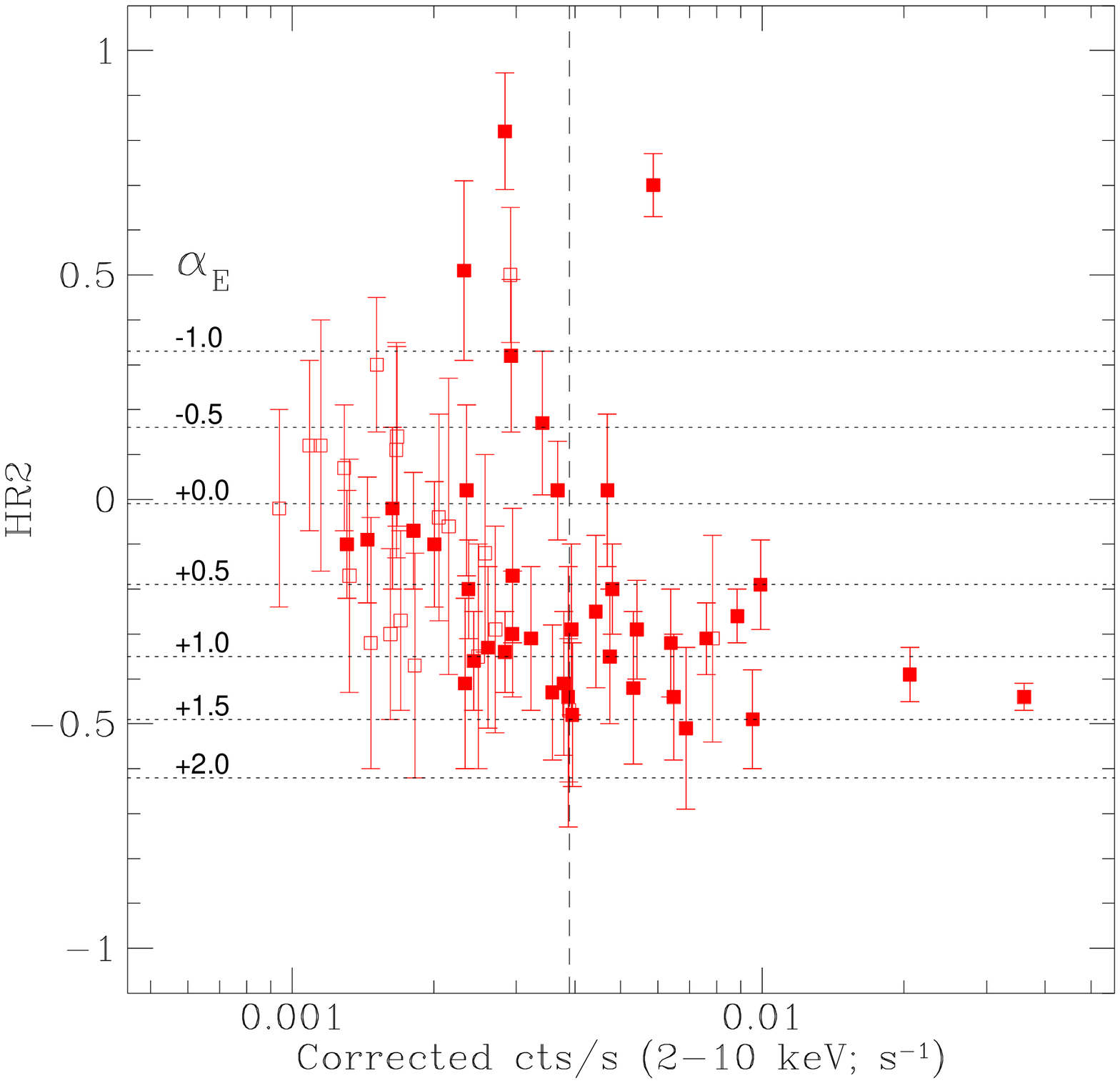}
\caption{}
\end{figure}

\clearpage
\begin{figure}
\figurenum{2}
\plotone{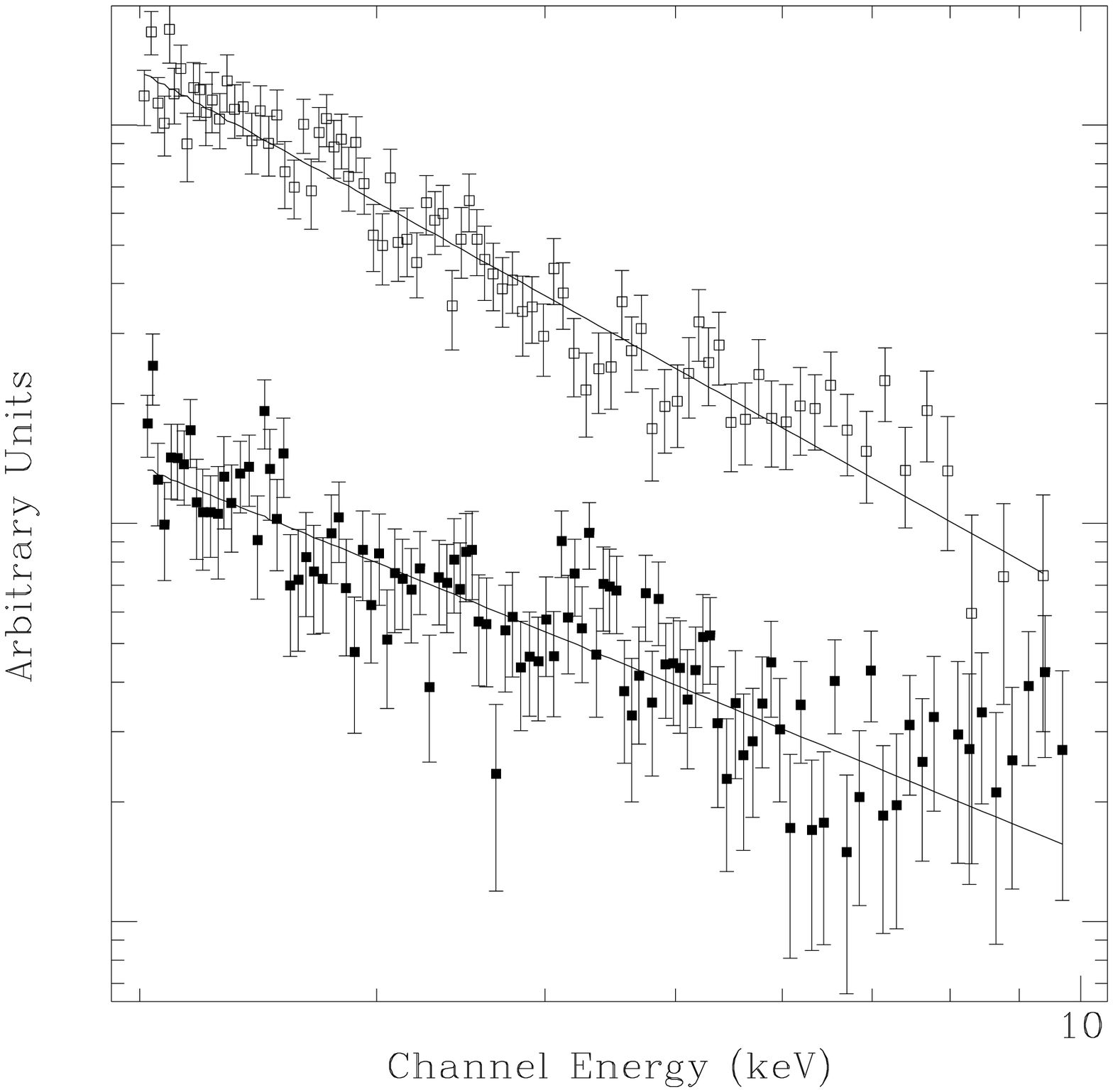}
\caption{}
\end{figure}

\clearpage
\begin{figure}
\figurenum{3}
\plotone{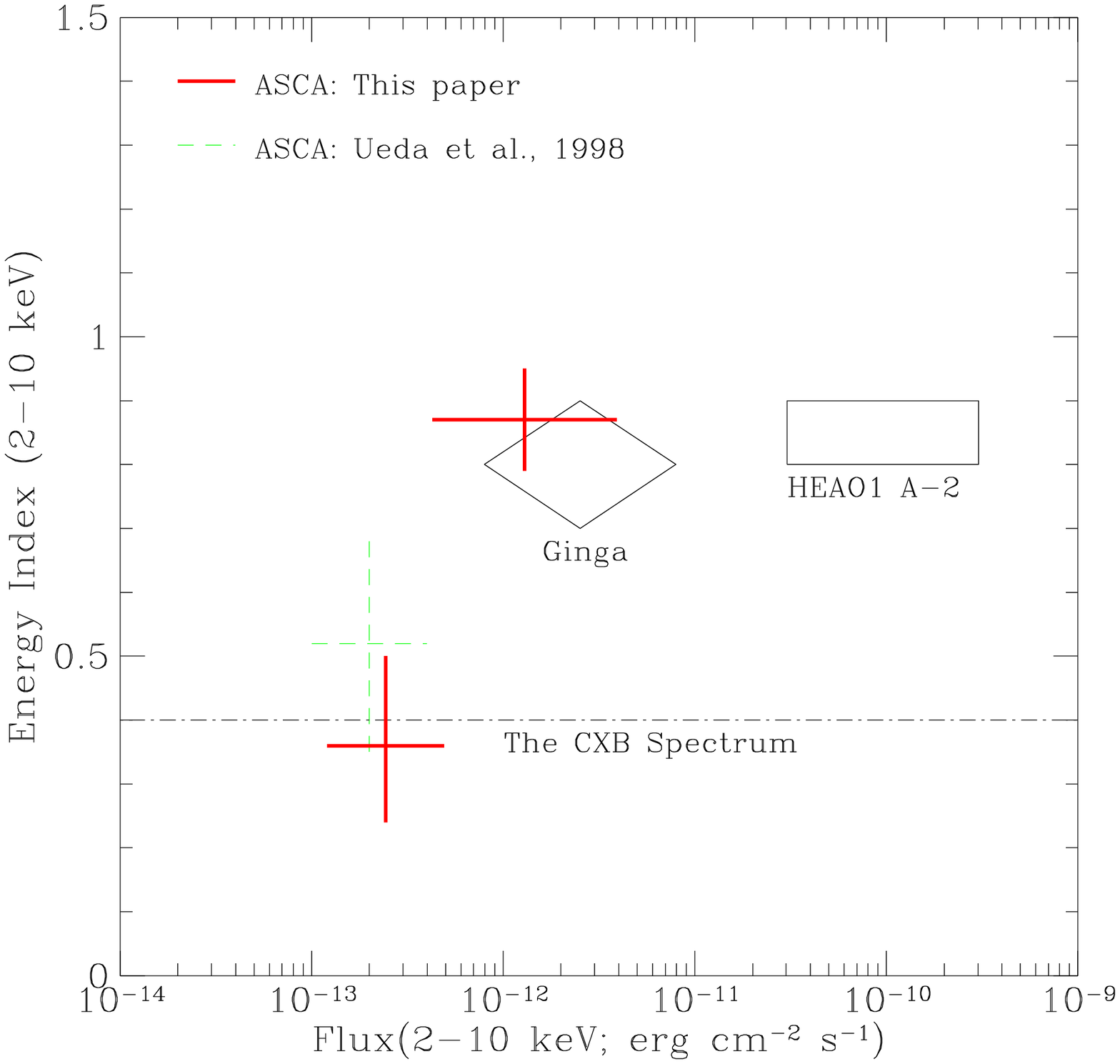}
\caption{}
\end{figure}

\clearpage
\begin{figure}
\figurenum{4}
\plotone{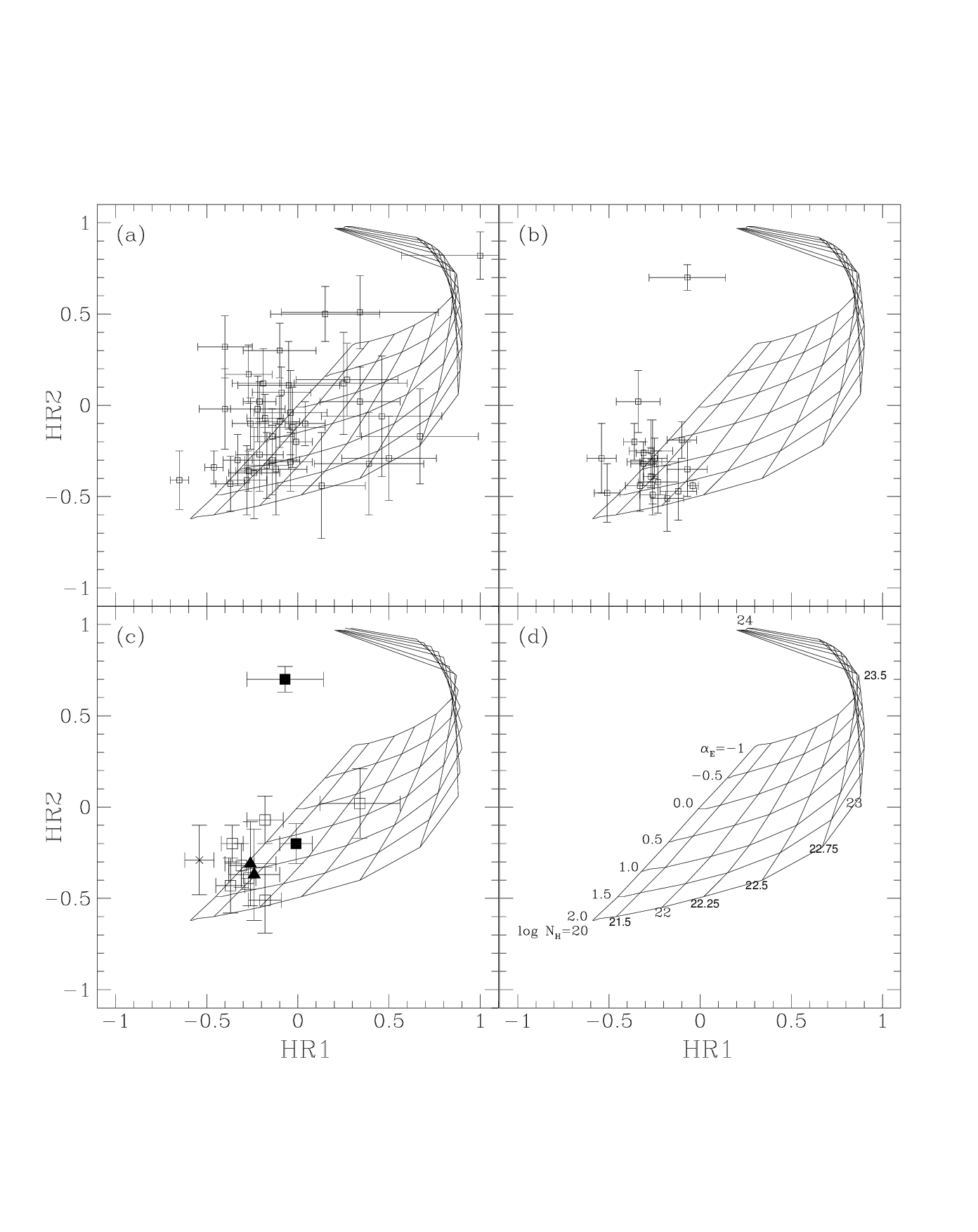}
\caption{}
\end{figure}

\clearpage
\begin{figure}
\figurenum{5}
\caption{This figure can be retrieved by ``anonymous ftp at
 {\it ftp.brera.mi.astro.it} in the directory "/pub/ASCA/paper2"}
\end{figure}

\clearpage
\begin{figure}
\figurenum{6}
\plotone{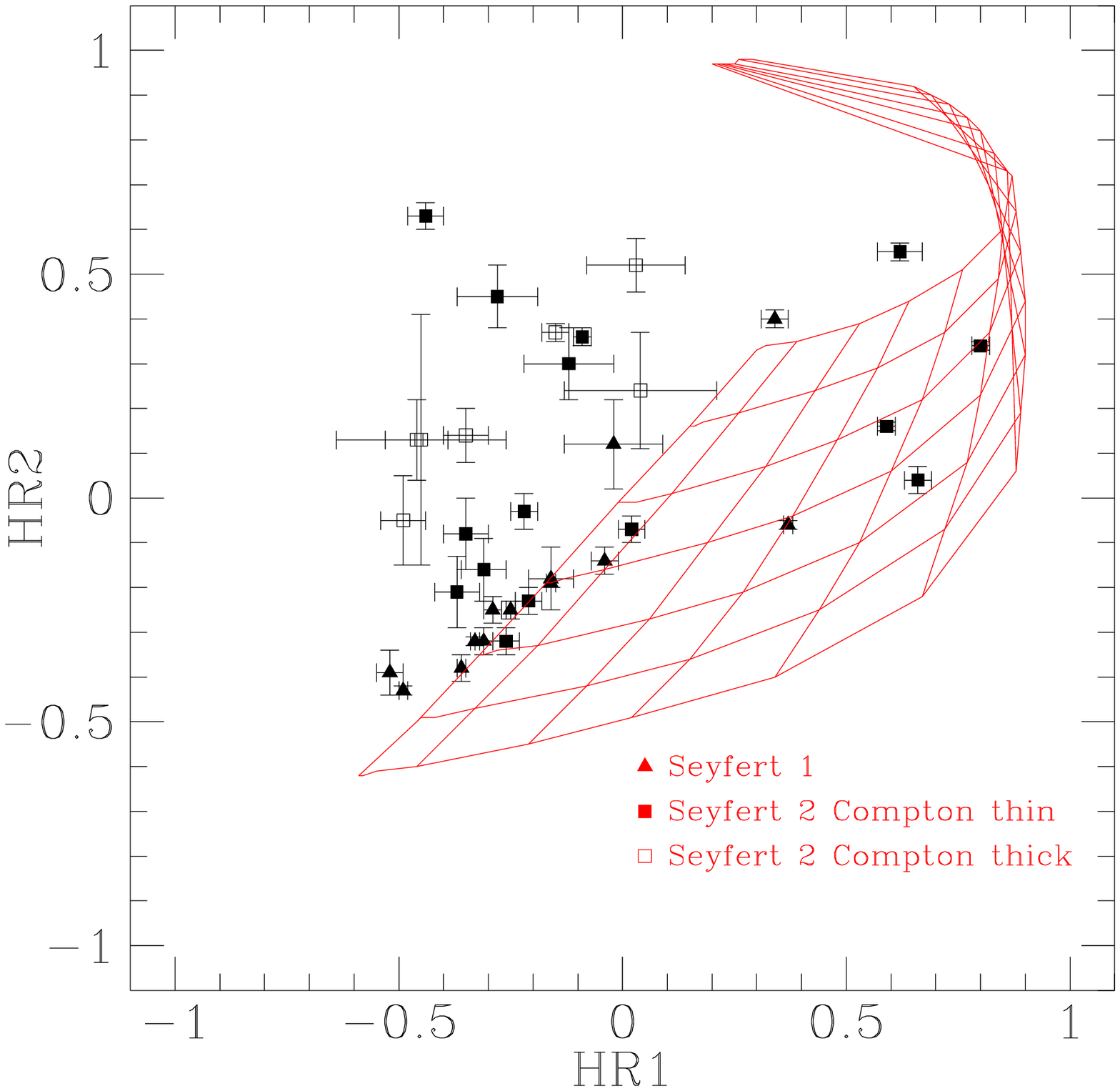}
\caption{}
\end{figure}

%

\end{document}